\documentclass[11pt,fleqn]{article}
\usepackage{psfig}
\begin{document}

\title{Population Inversion, Negative Temperature,\\ and Quantum Degeneracies}
\author{Zotin K.-H. Chu} % and Chen Qin$^{2}$}
\date{  %\address{ %\newline %}
%\address{
%}
WIPM, 30, West Xiao-Hong Shan, Wuhan 430071, PR China}
%
%$^{2}$ Department of Physics, Xinjiang Normal University, \\Urumqi
%830054,  China }
%\address{Institute of Physics Publishing, Dirac House, Temple Back, Bristol BS1 6BE, UK}
\maketitle
\begin{abstract}
We revisit the basic principle for lasing : Population inversion
which is nevertheless closely linked to the negative temperature
state in non-equilibrium thermodynamics. With the introduction of
quantum degeneracies, we also illustrate their relationship with
the lasing via the tuning of population inversion.
\end{abstract}
\doublerulesep=6.5mm    %\parskip=12 pt
\baselineskip=6.5mm  % %
\subsection*{Introduction} Atomic and molecular physics arc
certainly the oldest subfields of quantum physics and everybody
knows the major role that they played in the early 1920s when the
first principles of quantum mechanics were elaborated. After this
{\it golden era}, many physicists considered research in these
subfields essentially complete and in fact, for several decades,
activity in atomic and molecular physics decreased steadily in
comparison with the large effort directed towards nuclear and
high-energy physics. The advent of lasers (LASER means : Light
Amplification by Stimulated Emission of Radiation), and more
precisely of tunable lasers, changed this situation and since the
early 1970s, atomic and molecular physics, gradually intermingling
with laser physics, has become an area where more and more
activity is contributing substantially to the understanding of
many phenomena.
\newline
There has been a gradual change in the use of lasers  in the
teaching as well as research laboratory. Increasingly, the laser
is not being used for spectroscopy, but as a tool [1]. However,
the art of making a laser operate is to work out how to get
population inversion for the relevant transition. Once we have
population inversion, we have a mechanism for generating gain in
the laser medium. \newline
Thermodynamic arguments tell us, in addition to the black-body law
of radiation, that the interaction between electromagnetic waves
and matter at any temperature cannot produce amplification, for
radiation at the temperature of matter cannot be made more intense
by interaction of the two without violating the second law [2].
%\newline
%
\subsection*{Two-Level System}
We now wish to study how we can use stimulated emission to make a
light amplifier. In a gas of atoms in thermal equilibrium, the
population of the lower level ($N_1$) will always be greater than
the population of the upper level ($N_2$). (Please see Eq. (2)
below). Therefore, if a light beam is incident on the medium (cf.
Fig. 1), there will always be more upward transitions due to
absorption than downward transitions due to stimulated emission
(as the absorption is proportional to $N_1$ and the stimulated
emission is proportional to $N_2$). Hence there will be net
absorption, and the intensity of the beam will diminish on
progressing through the medium. To amplify the beam, we require
that the rate of stimulated emission transitions exceeds the rate
of absorption. If the light beam is sufficiently intense that we
can ignore spontaneous emission1 and the levels are
non-degenerate, this implies that $N_2$ must exceed $N_1$. This is
a highly non-equilibrium situation, and is called {\it population
inversion}.
\newline Inspection of Eq. (2) below implies that population inversion
corresponds to negative temperatures! This is not as ridiculous as
it sounds, because the atoms are not in thermal equilibrium. Once
we have population inversion, we have a mechanism for generating
gain in the laser medium. The art of making a laser operate is to
work out how to get population inversion for the relevant
transition.  \newline
The rate of change of electromagnetic energy confined in a region
where it interacts with a group of particles must, from Einstein's
work, have the form
\begin{equation}
 B_{12} N_1 u(\nu)=A_{21} N_2+ B_{21} N_2 u(\nu)
\end{equation}
where $N_1$ and $N_2$ are the numbers of molecules in the lower
and upper of two quantum states. In thermal equilibrium the ratio
of $N_2$ to $N_1$ at temperature $T$ is given by Boltzmann's law :
\begin{equation}
 \frac{N_2}{N_1}=\frac{g_2}{g_1} e^{-\hbar \nu/k_B T}
\end{equation}
where $g_2$, $g_1$ are the degeneracies of level 2,1 respectively,
and $\hbar \nu=E_2-E_1$ (if the energy of thermal motion is
sufficient : $k_B T > E_2 -E_1$, then a part of particles are
thrown into the upper level). \newline In fact, we may have the
following case : e.g., a light quantum may be absorbed by the
medium: and, in this case an absorption is produced. The
difference in energy between the upper and lower levels is equal
to the quantum energy. This process is connected with the decrease
in energy of the electromagnetic field and is called {\it
resonance absorption} [3].\newline On the other hand, we may also
have : Under the influence of a quantum, a particle may be
transferred from the upper level to the lower level. Such a
transfer will be accompanied by the emission of a light quantum
identical in frequency, direction of propagation and polarization
to the quantum which produced the emission. This process is
connected with an increase of the field energy and is called {\it
stimulated emission}. \newline
 We can
easily understand that, from above expression, we have, once both
of the two levels are non-degenerate,
\begin{equation}
 \frac{N_2}{N_1} \le 1 \hspace*{12mm} \mbox{if} \hspace*{6mm} T
 \ge 0.
\end{equation}
%------------------------
It is just this property of high ordering of a system with
negative temperature which makes it possible to produce
high-coherent emission in quantum oscillators, to produce high
sensitive quantum amplifiers, and to separate the energy stored in
the state with negative temperature in a very short time, of the
order of the reciprocal of the emission frequency [2].\newline
Meanwhile from the expression of Eq. (2), if both of the two
levels are degenerate and $e^{-\hbar \nu/k_B T}$is not the
dominated term at the right-hand-side of Eq. (2), we can still
have the population inversion (i.e., $N_2/N_1 >1$) via the tuning
of ratio of $g_2/g_1$ or degeneracies of level 2,1. The possible
illustration of equation (2) is shown in Fig. 2 where some cases
of $g_2/g_1$ and $\hbar \nu/k_B T$ were selected. Note that there
is no  population  inversion ($N_2/N_1 \le 1$) for $g_2/g_1 \sim
1$ and $\hbar \nu/k_B T >0.1$.
%\newline
\subsection*{Example : Argon Ion Laser}
To give an example for the latter situation, we shall consider
Argon which has 18 electrons with the configuration
1s$^2$2s$^2$2p$^6$3s$^2$3p$^6$. Argon atoms incorporated into a
discharge tube can be ionized by collisions with the electrons
Since there are six 4p levels as compared to only two 4s levels,
the statistics of the collisional process leaves three times as
many electrons in the 4p level than in the 4s level ($g_2/g_1=3$).
Hence we have population inversion. Moreover, cascade transitions
from higher excited states also facilitates the population
inversion mechanism. The lifetime of the 4p level is 10 ns, which
compares to the 1 ns lifetime of the 4s level. The partial
application of above reasoning leads to the Argon ion (normally
Ar$^+$) Lasers.
\subsection*{Higher-Level Systems}
Some lasers are classified as being three-level systems. The
standard example is ruby, which was the first laser ever produced.
The key difference between a three-level laser and a four-level
laser is that the lower laser level is the ground state. It is
much more difficult to obtain population inversion in three-level
lasers because the lower laser level initially has a very large
population. Let this population be $N_g$. By turning on the pump,
we excite $dN$ atoms to level 1, which then decay to level 2. Thus
the population of Level 2 will be $dN$, and the population of the
ground state will be ($N_g-dN$). Hence for population inversion we
require $dN > (N_g-dN)$, that is $dN
> N_g/2$. Therefore, in order to obtain population inversion we have
to pump more than half the atoms out of the ground state into the
upper laser level. This obviously requires a very large amount of
energy. This contrasts with the four-level lasers in which the
lower laser level is empty before the pumping process starts, and
much less energy is required to reach threshold. Despite the fact
that the threshold for population inversion is very high in a
three-level system, they can be quite efficient once this
threshold is overcome. Ruby lasers pumped by bright flash lamps
actually give very high output pulse energies. However, they only
work in pulsed mode. Continuous lasers tend to be made using
four-level systems.
\subsection*{Conclusion}
In this short paper by using the energy-level diagram for atoms
with two energy levels we revisit the population inversion that is
the fundamental principle for lasing and although it is closely
related to the negative temperature state in non-equilibrium
thermodynamics. We also illustrate the role of quantum
degeneracies or their relationship with the lasing via the tuning
of population inversion.
%------------------------
%\appendix

%\section*{References}

\newpage
%\subsubsection*{Acknowledgements}
%{\small  }
\psfig{file=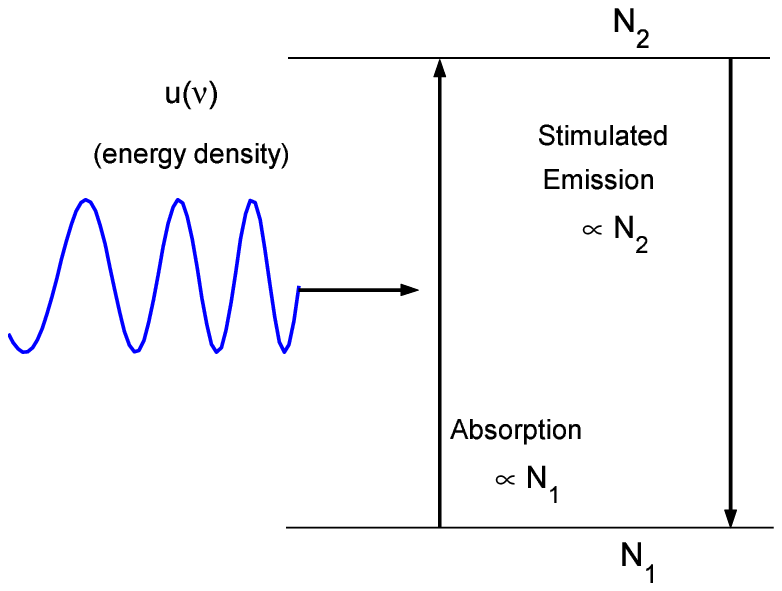,bbllx=1cm,bblly=18.5cm,bburx=14cm,bbury=25cm,rheight=6cm,rwidth=6cm,clip=}

\vspace{2mm}
\begin{figure}[h]
\hspace*{6mm} Fig. 1 \hspace*{1mm} Schematic 2-level system for
absorption and stimulated \newline \hspace*{7mm}  emission
transitions (atoms with two energy levels).
%experimental set-up
%for investigating \newline \hspace*{7mm} hyperfine structures with
%optical pumping approach [2].
%plot for the regular %scattering and the orientational scattering.
%Plane waves propagate along the $X$-direction.
%Binary encounters of $U_1$ and $U_3$ and their
%\newline
%\hspace*{7mm}
%departures  after head-on collisions ($U_2$ and
%$U_4$). Number densities $N_i$ are associated to $U_i$.
\end{figure}

\newpage
%\subsubsection*{Acknowledgements}
%{\small  }
\psfig{file=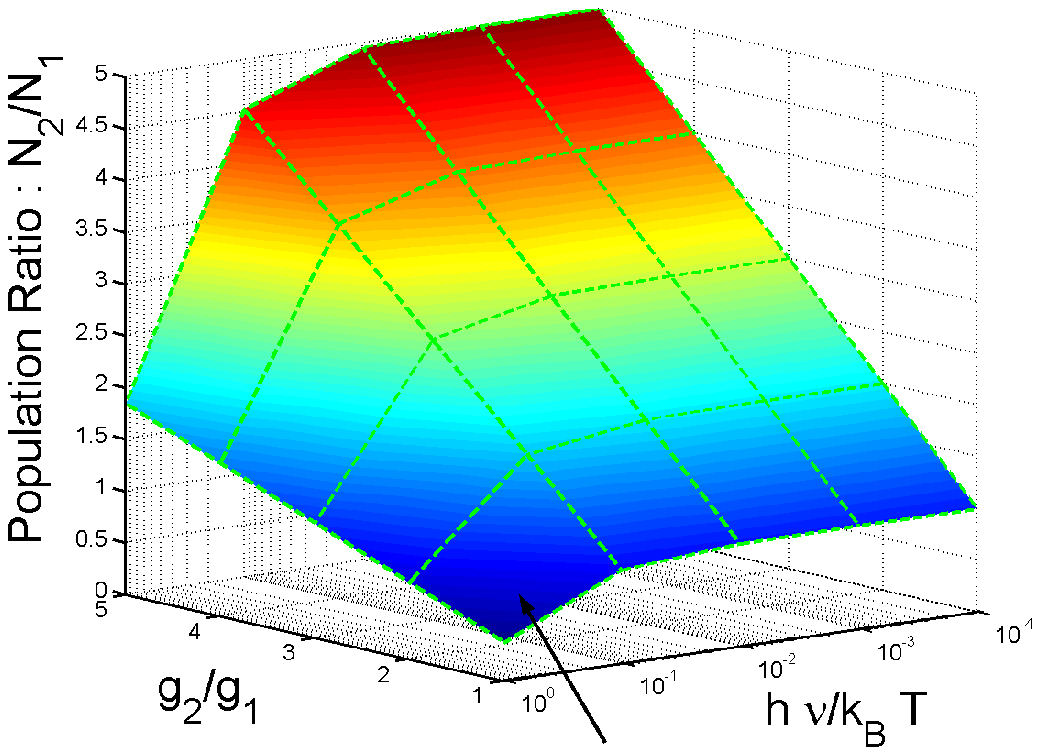,bbllx=0.1cm,bblly=15.6cm,bburx=14cm,bbury=28cm,rheight=12cm,rwidth=10cm,clip=}
\vspace{2mm}
\begin{figure}[h]
\hspace*{6mm} Fig. 2 \hspace*{1mm} Illustration of population
inversion ($N_2/N_1>1$, cf. Eq. (2)) \newline \hspace*{8mm} for
some cases  of degeneracies ($g_2/g_1$) and $\hbar \nu/k_B T$.
There is no \newline \hspace*{8mm} population  inversion ($N_2/N_1
\le 1$) for $g_2/g_1 \sim 1$ and $\hbar \nu/k_B T >0.1$ \newline
\hspace*{8mm}(the lower region where the (black) arrow points
out).
%Schematic hyperfine multiplet %corresponding to $D_2$ transition.
\end{figure}
%
%\newpage
%
%
\end{document}